\documentclass[journal,article,submit,moreauthors,pdftex]{Definitions/mdpi}

\usepackage{listings,xcolor}
\usepackage{rotating}

\renewcommand{\journalname}{Technologies}
\renewcommand{\linenumbers}{}

\firstpage{1} 
\pubvolume{1}
\issuenum{1}
\articlenumber{0}
\pubyear{2020}
\copyrightyear{2020}
\history{Received: date; Accepted: date; Published: date}

\usepackage{todonotes}

\Title{Enhanced Bug Prediction in JavaScript Programs\\with Hybrid Call-Graph Based Invocation Metrics}

\Author{Gábor Antal $^{1}$, Zoltán Tóth $^{1}$, Péter Hegedűs $^{1, 2}$* and Rudolf Ferenc $^{1}$}

\AuthorNames{Gábor Antal, Zoltán Tóth, Péter Hegedűs and Rudolf Ferenc}

\address{%
$^{1}$ \quad University of Szeged; \{antal|zizo|hpeter|ferenc\}@inf.u-szeged.hu\\
$^{2}$ \quad MTA-SZTE Research Group on Artificial Intelligence; hpeter@inf.u-szeged.hu}

\corres{Correspondence: hpeter@inf.u-szeged.hu}

\abstract{Bug prediction aims at finding source code elements in a software system that are likely to contain defects.
Being aware of the most error-prone parts of the program, one can efficiently allocate the limited amount of testing and code review resources.
Therefore, bug prediction can support software maintenance and evolution to a great extent.
In this paper, we propose a function level JavaScript bug prediction model based on static source code metrics with the addition of a hybrid (static and dynamic) code analysis based metrics of the number of incoming and outgoing function calls (HNII and HNOI).
Our motivation for this is that JavaScript is a highly dynamic scripting language for which static code analysis might be very imprecise, therefore using purely static source code features for bug prediction might not be enough.
Based on a study where we extracted 824 buggy and 1943 non-buggy functions from the publicly available BugsJS dataset for the ESLint JavaScript project, we can confirm the positive impact of hybrid code metrics on the prediction performance of the ML models.
Depending on the ML algorithm, applied hyper-parameters, and target measure we consider, hybrid invocation metrics bring a 2-10\% increase in model performances (i.e., precision, recall, F-measure).
Interestingly, replacing static NOI and NII metrics with their hybrid counterparts HNOI and HNII in itself improve model performances, however, using them all together yields the best results.}


\begin{document}

\section{Introduction}\label{sec:intro}

Bug prediction aims at finding source code elements in a software system that are likely to contain defects.
Being aware of the most error-prone parts of the program, one can efficiently allocate the limited amount of testing and code review resources.
Therefore, bug prediction can support software maintenance and evolution to a great extent.
However, practical adoption of such prediction models always depends on their real-world performance and the level of disturbing misclassification (i.e., false-positive hits) they produce.
Despite the relative maturity of the bug prediction research area, the practical utilization of the state-of-the-art models is still very low due to the reasons mentioned above.

Bug prediction models can use a diverse set of features to build effective prediction models.
The most common types of such features are static source code metrics~\cite{gray2013software, li2011mining, gray2009using, ferzund2008analysing}, process metrics~\cite{hata2012bug, shivaji2012reducing, madeyski2015process}, natural language features~\cite{binkley2009increasing, haiduc2016use}, and their combination~\cite{alshehri2018applying, di2018developer, goyal2013impact}.
All these metrics proved to be useful in different contexts, but the performance of these models may vary based on, for example, the language of the project, the composition of the project team, or the domain of the software product.
We need further studies to understand better how and when these models work best in certain situations.
Additionally, we can refine source code metrics by using static and dynamic analysis in combination, which has a yet unknown impact on the performance of bug prediction models.

In this paper, we propose a function level JavaScript bug prediction model based on static source code metrics with the addition of a hybrid (static and dynamic) code analysis based metrics of the number of incoming and outgoing function calls.
JavaScript is a highly dynamic scripting language for which static code analysis might be very imprecise.
Although static source code metrics proved to be very efficient in bug prediction, the imprecision due to the lack of dynamic information might affect the bug prediction models using on them.

To support the hybrid code analysis of JavaScript programs, we created a hybrid call-graph framework, called \emph{hcg-js-framework}\footnote{\url{https://github.com/sed-szeged/hcg-js-framework}} that can extract call-graph information of JavaScript functions using both static and dynamic analysis.
Based on the hybrid call-graph results of the ESLint\footnote{\url{https://eslint.org/}} JavaScript project, which we used as a subject system for bug prediction, we refined the Number of Incoming Invocations (NII) and Number of Outgoing Invocations (NOI) metrics. We added them to a set of common static source code metrics to form the predictor features in a training dataset consisting of 824 buggy and 1943 non-buggy functions extracted from the publicly available BugsJS~\cite{2019-Gyimesi-ICST} bug dataset\footnote{\url{https://github.com/BugsJS}}.
These invocation metrics are typically very imprecise in JavaScript calculated based only on static analysis, as lots of calls happen dynamically, like higher-order function calls, changes in prototypes, or executing the \emph{eval()} function, which is impossible to capture statically.
We analyzed the impact of these additional hybrid source code metrics on the function-level bug prediction models trained on this dataset.

We found that using invocation metrics calculated by a hybrid code analysis as bug prediction features consistently improves the performance of the ML prediction models.
Depending on the ML algorithm, applied hyper-parameters, and target measure we consider, hybrid invocation metrics bring a 2-10\% increase in model performances (i.e., precision, recall, F-measure).
Interestingly, even though replacing static NOI and NII metrics with their hybrid counterparts HNOI and HNII in itself improve model performances, keeping them all together yields the best results.
It implicates that hybrid call metrics indeed add some complementary information to bug prediction.

The rest of the paper is structured as follows.
In Section \ref{sec:related} we overview the JavaScript call-graph related literature and their usage for refining static source code metrics.
We summarize our methodology for collecting ESLint bugs, mapping them to functions, extracting hybrid call-graphs, and assembling the training dataset in Section \ref{sec:method}.
Section \ref{sec:results} contains the results of comparing bug prediction models using only static, only hybrid, or both static and hybrid metrics as features for machine learning models.
We enlist the possible threats to our work in Section \ref{sec:threats} and conclude the paper in Section \ref{sec:conclusion}.

\section{Related Work}\label{sec:related}

Using call-graphs for source code and program analysis is a well-established and mature technique; the first papers dealing with call-graphs date back to the 1970's~\cite{allen1974interprocedural, ryder1979constructing, Graham:1982:GCG:872726.806987}.
Call-graphs can be divided into two subgroups based on the method used to construct them: dynamic~\cite{xie2002empirical} and static~\cite{murphy1998empirical}.

Dynamic call-graphs can be obtained by the actual run of the program.
During the run, several runtime information is collected about the interprocedural flow~\cite{eichinger2010localizing}.
Techniques such as instrumenting the source code can be used for dynamic call-graph creation~\cite{Graham:1982:GCG:872726.806987, dmitriev2004profiling}.

In contrast, there is no need to run the program in the case of static call-graphs, as it is produced by a static analyzer which analyzes the source code of software without actually running it~\cite{Graham:1982:GCG:872726.806987}.
On the other hand, static call-graphs might include false edges (calls) since a static analyzer identifies several possible calls between functions that are not feasible in the actual run of a program; or they might miss real edges.
Static call-graphs can be constructed in almost any case from the source code, even if the code itself is not runnable.

Different analysis techniques are often combined to obtain a hybrid solution, which guarantees a more precise call-graph, thus a more precise analysis~\cite{eisenbarth2001aiding}.

With the spread of scripting languages such as Python and JavaScript, the need for analyzing programs written in these languages also increased~\cite{feldthaus_acg}.
However, constructing precise static call-graphs for dynamic scripting languages is a very hard task that is not fully solved yet~\cite{yu2019empirical}.
The \emph{eval()}, \emph{apply()}, and \emph{bind()} constructions of the languages make it especially hard to analyze the code statically.
There are several approaches to construct such static call-graphs for JavaScript with varying success~\cite{feldthaus_acg, bolin2010closure, fink2012wala, antal2018static}.
However, the most reliable method is to use dynamic approaches to detect such call edges.
We decided to use both dynamic and static analysis to ensure better precision even though it increases the analysis time, and the code should be in a runnable state due to the dynamic analysis.

Wei and Ryder presented blended taint analysis for JavaScript, which uses a combined static-dynamic analysis~\cite{wei2013practical}.
By applying dynamic analysis, they could collect information for even those situations that are hard to analyze statically.
Dynamic results (execution traces) are propagated to a static infrastructure, which embeds a call-graph builder as well.
This call-graph builder module makes use of the dynamically identified calls.
However, in the case of pure static analysis, they wrapped the WALA tool~\footnote{\url{https://github.com/wala/WALA}} to construct a static call-graph.
As previously said, our approach works similarly and also supports additional call graph builder tools to be included in the flow of the analysis.

Feldthaus et al. presented an approximation method to construct a call-graph~\cite{feldthaus_acg} by which a scalable JavaScript IDE support could be guaranteed.
We used a static call-graph builder tool in our toolchain, which is based on this approximation method.\footnote{\url{https://github.com/Persper/js-callgraph}}
Additional static JavaScript call-graph building algorithms were evaluated by Dijkstra~\cite{dijkstra2014evaluation}.
Madsen et al. focused on the problems induced by libraries used in the project~\cite{madsen2013practical}.
They used pointer analysis and a novel ``use analysis'' to enhance scalability and precision.

There are also works intending to create a framework for comparing call-graph construction algorithms~\cite{lhotak2007comparing, ali2012application}.
However, these are done for algorithms written in Java and~C.
Call-graphs are often used for preliminary analysis to determine whether an optimization can be done on the code or not.
Unfortunately, as they are specific to Java and C, we could not use these frameworks for our paper.

Clustering call-graphs can have advantages in malware classification~\cite{kinable2011malware}, they can help localizing software faults~\cite{eichinger2008mining}, not to mention the usefulness of call-graphs in debugging~\cite{rao2013debugging}.


Musco et al.~\cite{Musco2016Large} used four types of call-graphs to predict the software elements that are likely to be impacted by a change in the software.
However, they used mutation testing to assess the impact of a change in the source code.
The same methodology could have been used but with a slight change: 
instead of using an arbitrary change, it can be a vulnerability introducing or a vulnerability mitigating change.

Nuthan Munaiah and Andrew Meneely ~\cite{Munaiah2016BeyondTA} introduced two novel attack surface metrics with their approach, which are the "Proximity" and "Risky Walks" metrics.
Both of them are defined by the call-graph representation of the program.
Their empirical study proved that using their metrics to build a prediction model can help to predict more accurately as their metrics are statistically significantly
associated with the vulnerable functions.

Nguyen et al.~\cite{nguyen2010predicting} proposed a model to predict vulnerable components based on a metric set generated from the component dependency graph of a software.

Cheng et al.~\cite{cheng2019static} presented a new approach to detect control-flow-related vulnerabilities called VGDetector.
They applied a recent graph convolutional network to embed code fragments in a compact representation (while the representation still preserves the high-level control-flow information).

Neuhaus et al.~\cite{neuhaus2007predicting} presented a fully automatic way to map vulnerabilities to software components and a tool called Vulture that can automatically build predictors to predict vulnerabilities in a new component.
They identified that imports and function calls have an impact on whether a component vulnerable or not.
They also made an evaluation of Mozilla's codebase that showed that their approach is accurate.

Lee at al.~\cite{Lee2010Detecting} proposed a new approach to generate semantic signatures from programs to detect malware.
They extracted the call-graph of the API call sequence that would be generated by malware, called code graph.
This graph is used for the semantic signature.
They used semantic signatures to detect malware even if the malware is obfuscated or the malware slightly differs from its previous versions (these are the main reasons why a commercial anti-virus does not detect them as malware).

As these previous studies show, the advantage of call-graphs is present in predicting vulnerabilities in software systems.
We did not narrow down the type of defects.
Our approach is generally applicable to arbitrary bug prediction.

The most similar to our study is possibly the work of
Punia et al.~\cite{punia2014evaluation} who presented a call-graph based approach to predict and detect defects in a given program.
They also defined call-graph based metrics such as \textit{Fan In, Fan Out, Call-Graph Based Ranking (CGBR) and Information Flow Complexity (IFC)}.
They investigated the correlation between their metrics and several types of bugs.
They proved the hypothesis that there is a correlation between call-graph based metrics and bugs in software design.
The authors performed their study in the Java domain; contrarily, we focused on JavaScript systems.
Besides J84, LMT, and SMO, we applied additional machine learning algorithms and also evaluated deep learning techniques to find potential bugs in the software.
As many papers, we also focused on different source code metrics; however, we adopted coupling metrics for the so-called hybrid call-graph.

\section{Methodology}\label{sec:method}

\begin{figure}[htbp]
\begin{center}
\includegraphics[width=\textwidth]{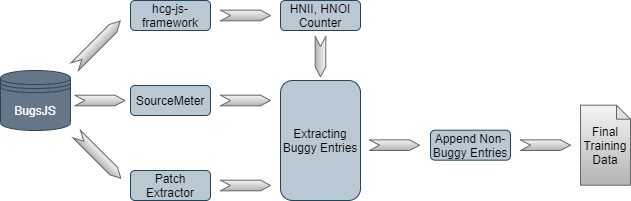}
\caption{Applied Toolchain}
\label{fig:toolchain}
\vspace{-20px}
\end{center}
\end{figure}

Our approach consists of numerous steps, which we present in detail in this section.
\figurename~\ref{fig:toolchain} shows the steps required to produce input for the machine learning algorithms.

\subsection{BugsJS Dataset}

BugsJS~\cite{2019-Gyimesi-ICST} is a bug dataset inspired by Defects4J~\cite{JustJE2014}; however, it provides bug related information for popular JavaScript-based projects instead of Java projects.
Currently, BugsJS includes bug information for ten projects that are actively maintained Node.js server-side programs hosted on GitHub.
Most importantly, BugsJS includes projects which adopt the Mocha testing framework; consequently, we can implement dynamic analysis experiments easier.

BugsJS stores the forks of the original repositories and extends them by adding tags for their custom commits in the form of:

\begin{itemize}
    \item \texttt{Bug-X}: The parent commit of the revision in which the bug was fixed (i.e., the buggy revision)
    \item \texttt{Bug-X-fix}: A revision (commit) containing only the production code changes (test code and documentation changes were excluded) introduced in order to fix the bug
\end{itemize}

\noindent, where \texttt{X} denotes a number associated with a given bug.
As out of the total 453 bugs, ESLint~\footnote{\url{https://github.com/eslint/eslint}} itself contains 333 bugs, we chose this project as input in our study.

\subsection{Hybrid Invocation Metrics Calculation}

\begin{figure}[htbp]
\begin{center}
\includegraphics[width=0.75\textwidth]{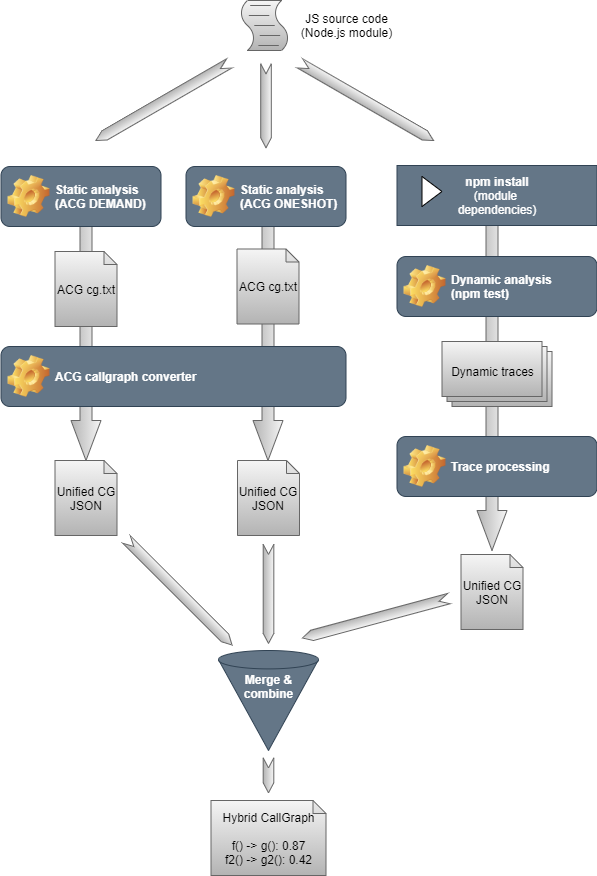}
\caption{Hybrid call-graph framework architecture}
\label{fig:hcg-framework}
\end{center}
\end{figure}

As a first step, we have to produce the so-called \textit{hybrid call-graphs} from which we can calculate the \textit{hybrid invocation metrics} (i.e., HNII and HNOI).
In order to understand what a hybrid call-graph is, let us consider \figurename~\ref{fig:hcg-framework}, which shows the details of the node "hcg-js-framework" presented earlier in \figurename~\ref{fig:toolchain}.

As can be seen, the input of the \textit{hcg-js-framework} is the JavaScript source code that we want to analyze, which can be either a Git repository or a local folder.
Then we analyze the source code with various static and dynamic tools.
Following the analyses, the framework converts all the tool-specific outputs to a unified JSON format.
Once we have the JSON files, the framework combines them into a merged JSON.
This merged JSON contains every node and edge (JavaScript function nodes and call edges between them) that either of the tools found.

After this step, we augment this merged JSON with confidence levels for the edges.
The confidence levels are calculated based on a manual evaluation of 600 out of 82,791 call edges found in 12 real-world Node.js modules.
We calculated the True Positive Rate for each tool intersection.
We estimate the confidence of a call edge with these rates.
For instance, if a call edge was found by tools A and B, and in the manually evaluated sample, there were ten edges found by only these tools, from which five turned out to be a valid call edge, we add confidence of 0.5 to all these edges.

\begin{figure}[htbp]
\begin{center}
\includegraphics[width=0.8\textwidth]{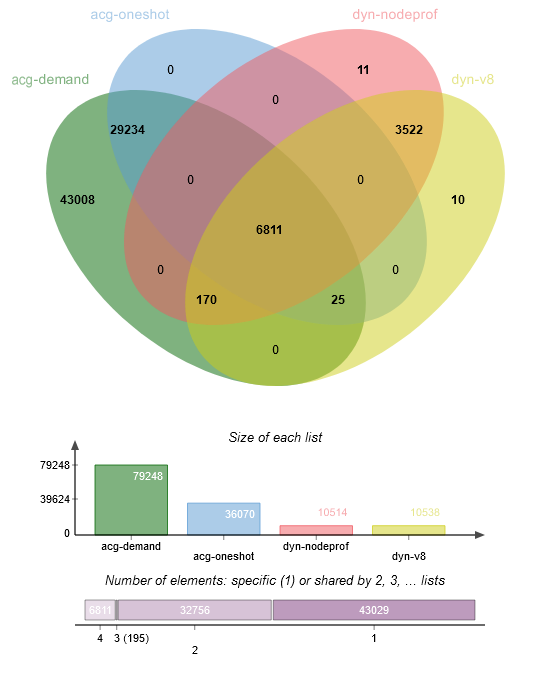}
\caption{Venn diagram of found edges}
\label{fig:venn}
\end{center}
\end{figure}

Figure~\ref{fig:venn} shows a Venn diagram of the call edges found in 12 Node.js modules. We have an evaluation ratio for each intersection, which the framework uses for edge confidence level estimation.

To sum it up, a \textit{hybrid call-graph} is a call-graph (produced by combining the results of both static and dynamic analysis) which associates a confidence factor to each call edge, which shows how likely an edge is valid (higher confidence means higher validity).

This hybrid call-graph is the input of the \textit{HNII, HNOI Counter} which is responsible for calculating the exact number of incoming and outgoing invocations (i.e., NII, NOI).
At this point, we have to specify the \textit{threshold} value, which defines the lower limit from which we consider a call edge as a valid call edge, thus contributing to the value of the number of incoming and outgoing invocations.
We considered four threshold values: \textit{0.00}, \textit{0.05}, \textit{0.20} and \textit{0.30}.
In the case of the first one, all edges are considered as possibly valid call edges, while the latter one only includes edges with a high confidence factor.
We name these two new metrics as HNII (Hybrid Number of Incoming Invocations) and HNOI (Hybrid Number of Outgoing Invocations) to differentiate them from the original static NII and NOI metrics.

The \textit{HNII, HNOI Counter} traverses all call edges and considers those edges where the confidence level is above the given threshold.
The edges fulfilling this threshold criteria contribute to the HNOI metric of the source node and the HNII metric of the target node.
As a result, the tool produces a JSON file as its output, which contains only the nodes (i.e., the JavaScript functions) with their corresponding HNII, HNOI metric values, and additional information about their position in the system, such as source file, line, and column.
Listing~\ref{lst:hnii-hnoi-counter} shows an example of a single node output.

\begin{lstlisting}[caption={Sample output from the HNII, HNOI Counter},label={lst:hnii-hnoi-counter}]
{
    "pos":"eslint/lib/ast-utils.js:169:25",
    "entry":false,
    "final":false,
    "hnii":1,
    "hnoi":3
}
\end{lstlisting}

\subsection{SourceMeter and Patch Extraction}\label{sec:osa-metrics}

Besides computing the HNII and HNOI metrics, a standard set of metrics is provided by a static source code analyzer named SourceMeter~\footnote{https://www.sourcemeter.com/}.
SourceMeter also takes JavaScript source code as input and outputs (amongst others) different CSV files for different source code elements (functions, methods, classes, files, system).
In this study, we used the resulting CSV file that contains function level entries, which captures static size metrics (LOC, LLOC, NOS), complexity metrics (McCC, NL), documentation metrics (CD, CLOC, DLOC), and traditional coupling metrics (NII, NOI) as well.
These metrics are calculated for all the 333 bugs in ESLint before and after the bug is fixed, which means 666 static analyses in total.

Similarly, we extracted the patches for these 333 bug fixing commits, which is done by \textit{Patch Extractor}.

\subsection{Composing Buggy Entries}

At this point, we have all the necessary inputs to combine them in one CSV, which contains the buggy entries with their static source code metrics extended with the HNII and HNOI metrics.
The core of the algorithm is the following.
We traverse all the bugs one-by-one.
For $bug_i$, we retrieved a set of entries from the $i^{th}$ static analysis results, which were touched by the fixing $patch_i$ (determined based on entry name and positional information) and extended these entries with the corresponding HNII and HNOI metric values.
We included the before-fix state (i.e., the buggy) for the touched JavaScript functions, and used the date of the latest bug to select non-buggy instances from that corresponding version of the code (i.e., Bug-79 fixed at 2018-03-21 17:23:34).
For non-buggy entries, we also extracted the corresponding HNII and HNOI values also from the latest buggy version.

\section{Results}\label{sec:results}

To calculate the HNOI and HNII metrics, one needs to apply a threshold to the call edges (to decide which edges to consider as valid) in the underlying hybrid (also called as fuzzy) call-graph produced by the hcg-js-framework (see Section \ref{sec:method}).
We calculated the metric values\footnote{All the data used in this study is available online~\cite{antal2020enhancedTrainingData}} with four different thresholds: 0, 0.05, 0.2, and 0.3.
Table~\ref{tab:thresh} shows the descriptive statistics of the metrics on our ESLint dataset.

\begin{table}[htbp]
  \centering
  \caption{Descriptive statistics of the HNII and HNOI metrics calculated using different thresholds}
    \begin{tabular}{lrrrrrr}
          & \multicolumn{3}{c}{\textbf{HNII}} & \multicolumn{3}{c}{\textbf{HNOI}} \\
    \textbf{Threshold} & \multicolumn{1}{l}{\textbf{Avg.}} & \multicolumn{1}{l}{\textbf{Median}} & \multicolumn{1}{l}{\textbf{Std.dev.}} & \multicolumn{1}{l}{\textbf{Avg.}} & \multicolumn{1}{l}{\textbf{Median}} & \multicolumn{1}{l}{\textbf{Std.dev.}} \\
    $>$ 0.00 & 7.026021 & 1     & 26.95583 & 5.341887 & 2     & 27.27586 \\
    $>$ 0.05 & 6.96133 & 1     & 26.95997 & 5.243224 & 2     & 26.91228 \\
    $>$ 0.20 & 0.840622 & 1     & 2.823739 & 1.018793 & 0     & 9.236607 \\
    $>$ 0.30 & 0.840622 & 1     & 2.823739 & 1.018793 & 0     & 9.236607 \\
    \end{tabular}%
  \label{tab:thresh}%
\end{table}%

As can be seen, thresholds 0.20 and above significantly reduces the number of considered edges for HNII and HNOI calculation.
We wanted to use as many of the extracted call edges as possible, so we selected to use the 0.00 threshold later on (i.e., we considered each edge in the fuzzy call-graph where the weight/confidence is greater or equal to zero).

We trained several models on the dataset with three different configurations for the features:
\begin{itemize}
    \item Purely static metrics ($S - 0\_00\_s.csv$): the dataset contains only the pure static source code metrics (i.e., original versions of NOI and NII plus all the provided metrics by SourceMeter, see Section \ref{sec:osa-metrics}).
    \item Static metrics with only hybrid NOI and NII versions ($H - 0\_00\_h.csv$): the dataset contains all the static metrics except NOI and NII, which are replaced by their hybrid counterparts (HNOII and HNII) calculated on the output of \emph{hcg-js-framework}.
    \item Both static and hybrid metrics ($S+H - 0\_00\_s+h.csv$): the dataset contains all the static metrics plus the hybrid counterparts of NOI and NII (HNOII and HNII) calculated on the output of \emph{hcg-js-framework}.
\end{itemize}

To have a robust understanding of the hybrid metrics' impact, we trained nine different machine learning models:
\begin{itemize}
    \item Logistic Regression Classifier -- Logistic regression is a statistical model that uses a logistic function to model a binary dependent variable (implemented by \verb|sklearn.linear_model.LogisticRegression|);
    \item Naive Bayes Classifier -- Naive Bayes classifier is a simple ``probabilistic classifier'' based on applying Bayes' theorem with strong (naïve) independence assumptions between the features (implemented by \verb|sklearn.naive_bayes.GaussianNB|);
    \item Decision Tree Classifier -- Decision Trees (DTs) are a non-parametric supervised learning method used for classification and regression, where the goal is to create a model that predicts the value of a target variable by learning simple decision rules inferred from the data features (implemented by \verb|sklearn.tree.DecisionTreeClassifier| an optimized version of the CART algorithm);
    \item Linear Regression Classifier -- Linear regression is a linear approach to modeling the relationship between a scalar response and one or more explanatory variables also known as dependent and independent variables (implemented by \verb|sklearn.linear_model.LinearRegression|);
    \item Standard DNN Classifier -- A deep neural network (DNN) is an artificial neural network (ANN) with multiple layers between the input and output layers (implemented using \verb|tensorflowf.layers.dense|);
    \item Customized DNN Classifier -- A custom version of the standard DNN implementing th early stopping mechanism, where we do not train the models for a fixed number of epochs, rather stop when there is no more reduction in the loss function (implemented using \verb|tensorflowf.layers.dense|);
    \item Support Vector Machine Classifier -- Support-vector machine (SVM) is a supervised learning model, which is a representation of the examples as points in space, mapped so that the examples of the separate categories are divided by a clear gap that is as wide as possible (implemented by \verb|sklearn.svm.SVC|);
    \item K Nearest Neighbors Classifier --  The k-nearest neighbors algorithm (k-NN) is a non-parametric method for classification and regression, where the input consists of the k closest training examples in the feature space (implemented by \verb|sklearn.neighbors.KNeighborsClassifier|);
    \item Random Forest Classifier -- Random forest is an ensemble learning method for classification, regression and other tasks that operates by constructing a multitude of decision trees at training time and outputting the class that is the mode of the classes (classification) or mean/average prediction (regression) of the individual trees (implemented by \verb|sklearn.ensemble.RandomForestClassifier|).
\end{itemize}
With the various hyper-parameters, it added up to a total of 36 configurations.
We executed all these 36 training tasks on all three feature sets, so we created 108 different ML models for comparison.
To cope with the highly imbalanced nature of the dataset (i.e., there are significantly more non-buggy functions than buggy ones), we applied a 50\% oversampling on the minority class.
We also standardized all the metric values to bring them to the same scale.
For the model training and evaluation, we used our open-source DeepWater Framework\footnote{\url{https://github.com/sed-inf-u-szeged/DeepWaterFramework}}~\cite{FERENC2020100551}, which contains the implementation of all the above algorithms.

\begin{figure}
\begin{center}
\includegraphics[width=\textwidth]{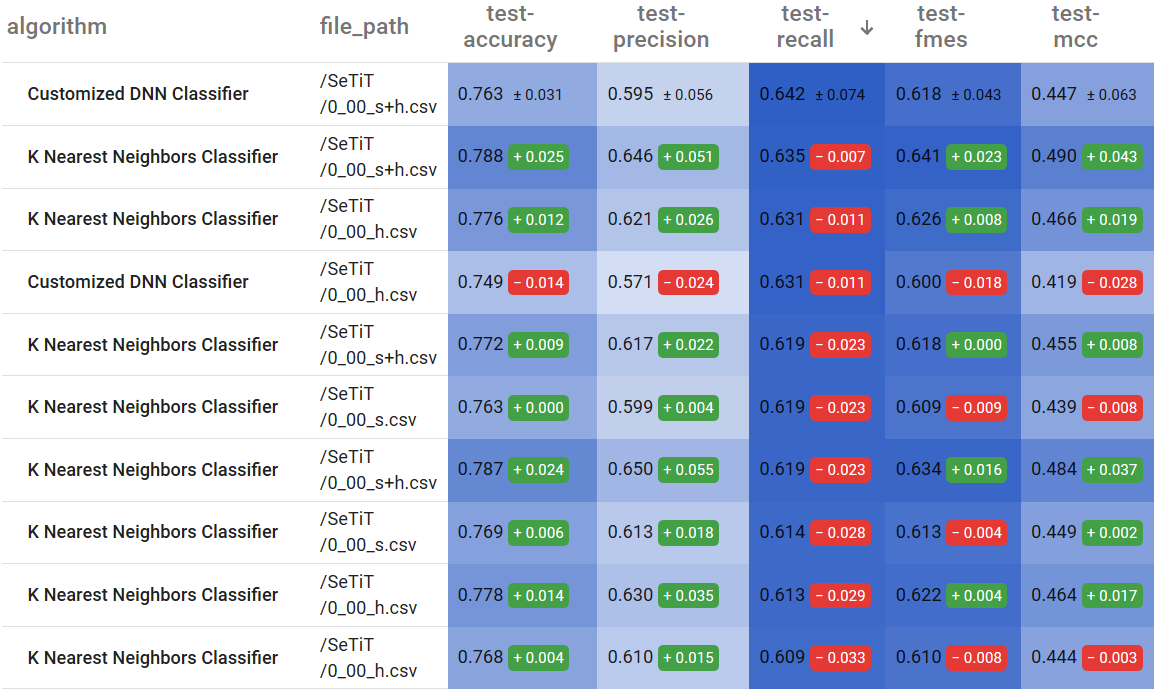}
\caption{Top 10 recall measures}
\label{fig:recall}
\end{center}
\end{figure}

To ensure that the results are robust against the chosen threshold, we trained the same 108 models with the threshold value of 0.30 for HNII and HNOI calculation.
We found that the differences among $S$, $H$, and $S+H$ feature sets became less, but the general tendency that $H$ and especially $S+H$ features achieved better results remained.
Therefore, in the rest of the paper, we can use the HNII and HNOI metrics calculated with the 0.00 threshold without the loss of generality.
In the remaining, we present our findings.

\subsection{The Best Performing Algorithms}

Figure~\ref{fig:recall} displays a heat-mapped table of the top 10 model results based on their recall values.
We ranked all 108 models, meaning that all three feature sets are on the same list.
We can measure recall with the following formula:
$$
    Recall = \frac{TP}{TP+FN}
$$,
where TP means True Positive samples, while FN means False Negatives.
As we can see, DNN (0.642) and KNN (0.635) models achieve the best recall values on the $S+H$ feature set.
The same models produce almost as high recall values (0.631) using only the $H$ feature set.
The best performing model on the $S$ feature set is KNN, with a significantly lower (0.619) recall value.
It shows that hybrid invocation metrics do increase the performance of ML models in terms of recall.
The best values are achieved by keeping both the original NOI and NII metrics and adding their hybrid counterparts HNOI and HNII, but using only the latter ones as substitutes for the static metrics still improves recall values.

\begin{figure}[htb!]
\begin{center}
\includegraphics[width=0.85\textwidth]{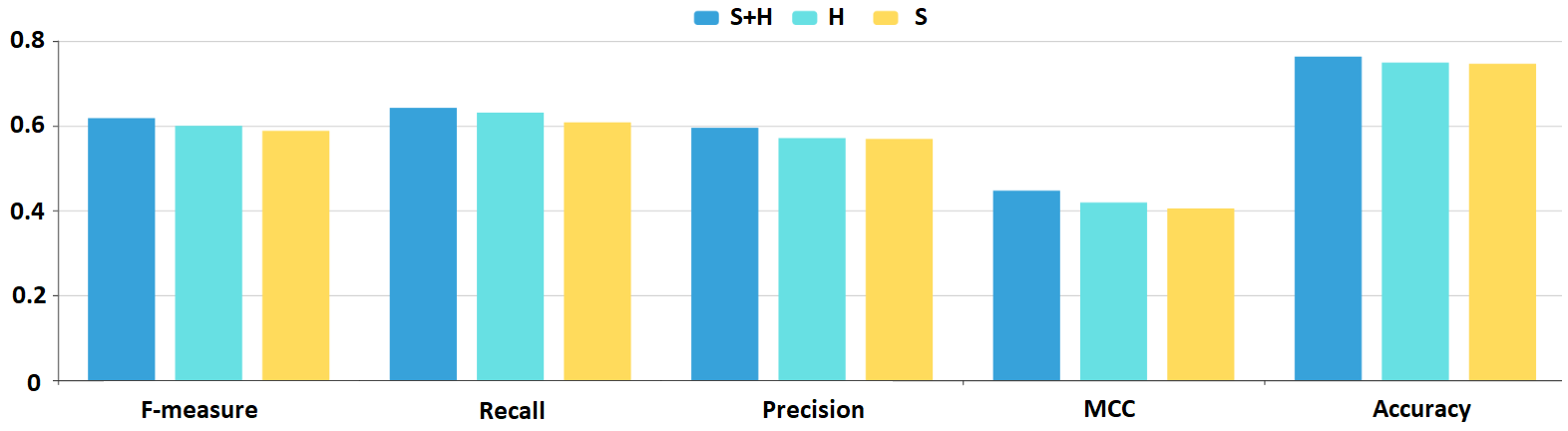}
\caption{Deep neural network}
\label{fig:dnn1}
\end{center}
\end{figure}

\begin{figure}[htb!]
\begin{center}
\vspace{-20px}
\includegraphics[width=0.85\textwidth]{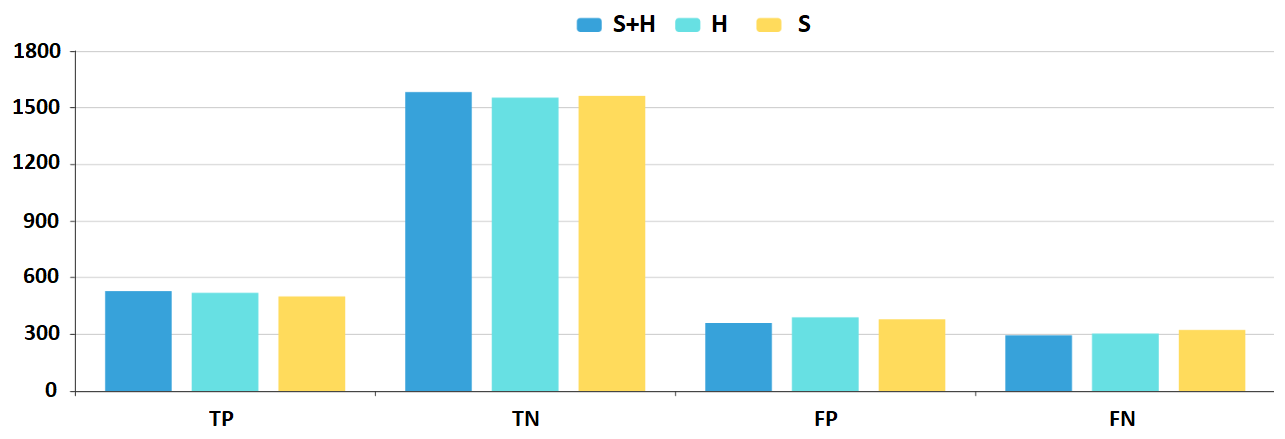}
\caption{Deep neural network}
\label{fig:dnn2}
\end{center}
\end{figure}

To visualize the difference in the various performance measures, we plotted a bar-chart (Figures~\ref{fig:dnn1} and ~\ref{fig:dnn2}) with the best DNN configurations (i.e., applying the set of hyper-parameters with which the model achieves the best performance) for all three feature sets.
Blue marks the results using the $S+H$ feature set, cyan the $H$ feature set, while yellow the $S$ feature set.
$S+H$ results are superior, while $H$ results are better than the $S$ results except for the False Positive and True Negative instances.
The chart shows that there is a constant 3-4\% improvement in all aspects of the DNN model results if we add the hybrid metrics to the feature sets.

\begin{figure}
\begin{center}
\includegraphics[width=\textwidth]{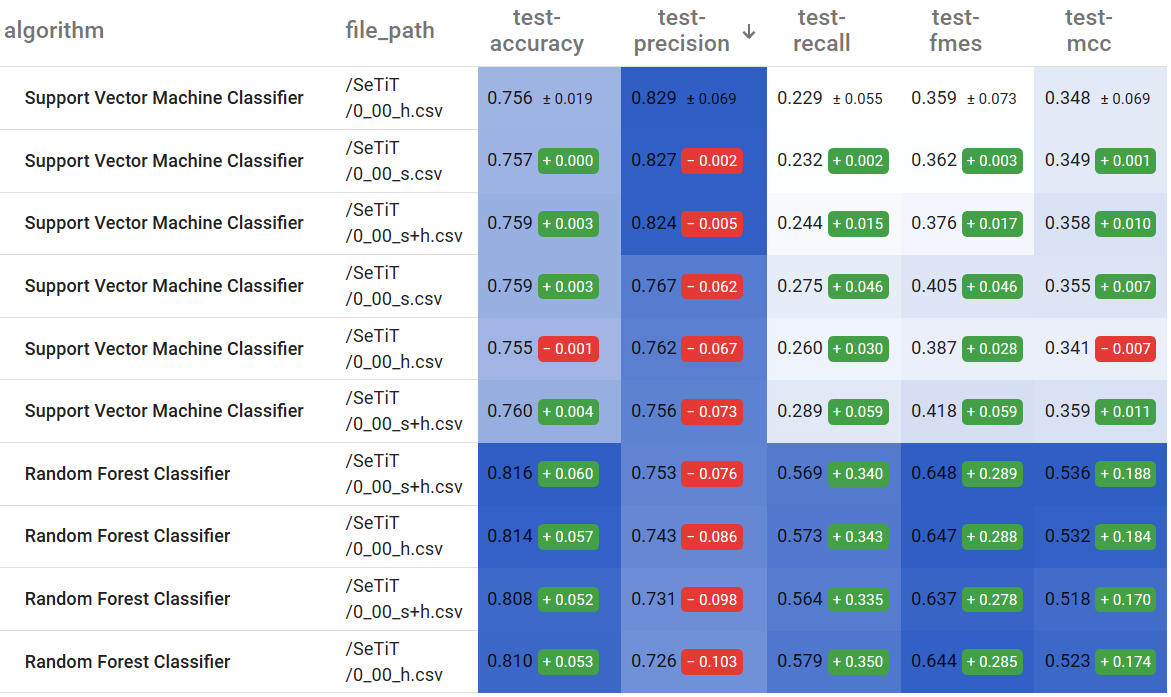}
\caption{Top 10 precision measures}
\label{fig:precision}
\end{center}
\end{figure}

Figure~\ref{fig:precision} displays a heat-mapped table of the top 10 model results based on their precision values.
We ranked all 108 models, meaning that all three feature sets are on the same list.
We can measure precision with the following formula:
$$
    Precision = \frac{TP}{TP+FP}
$$,
where TP means True Positive samples, while FP means False Positives.
As we can see, the SVM model (0.829) achieves the best precision values on the $H$ feature set.
Interestingly, SVM produces an almost as high precision value (0.827) using only the $S$ feature set as well.
Based on the $S+H$ feature set, SVM achieves a precision value of 0.824.
It shows that hybrid invocation metrics do increase the performance of ML models in terms of precision, but not as significantly as in the case of recall values.
Nonetheless, for other algorithms than SVM, the increase is more significant.

\begin{figure}[htbp]
\begin{center}
\includegraphics[width=0.85\textwidth]{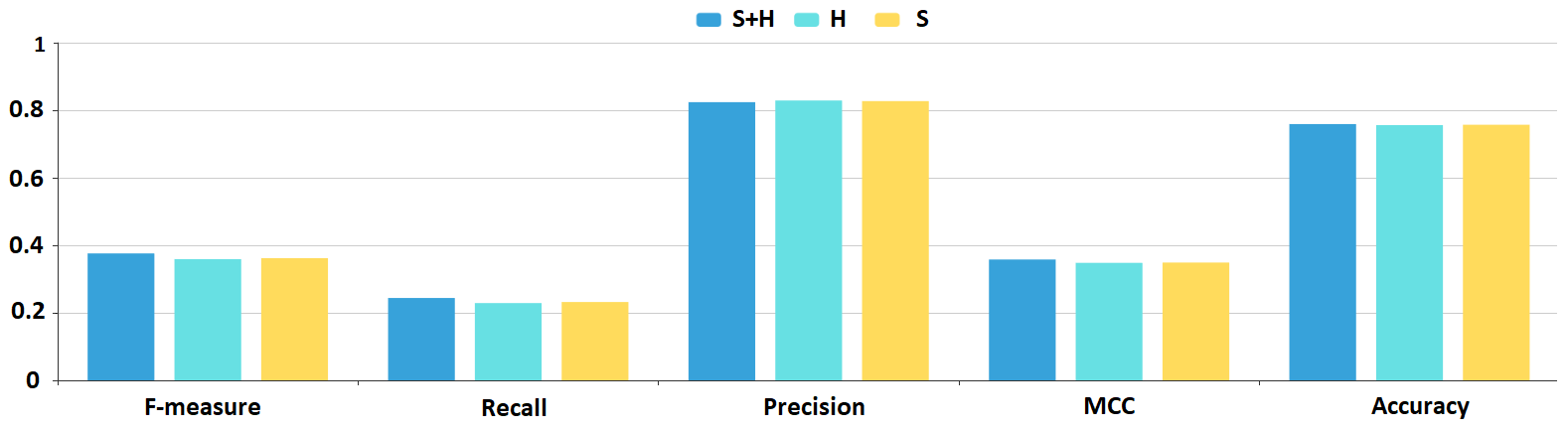}
\caption{SVM}
\label{fig:svm1}
\end{center}
\end{figure}

\begin{figure}[htbp]
\begin{center}
\vspace{-20px}
\includegraphics[width=0.85\textwidth]{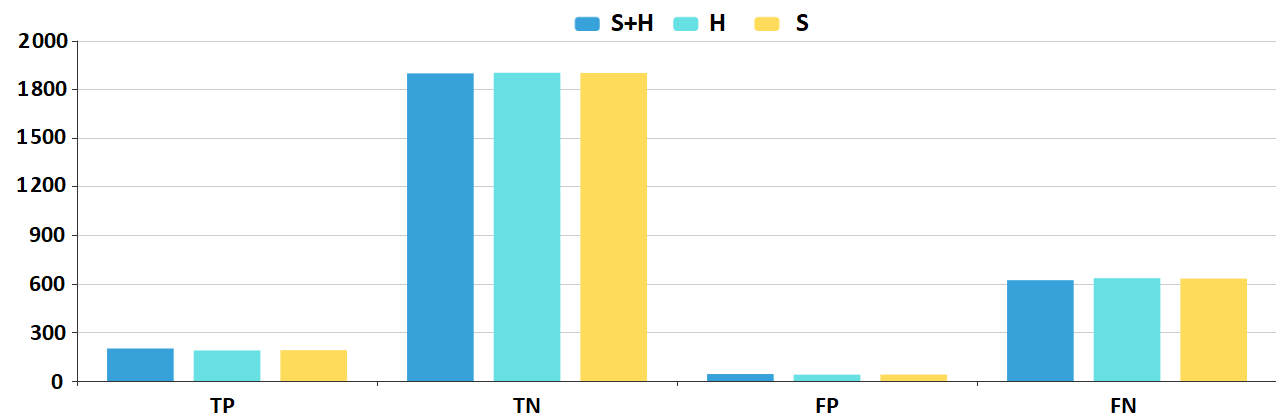}
\caption{SVM}
\label{fig:svm2}
\end{center}
\end{figure}

To visualize the difference in the various performance measures, we plotted a bar-chart (Figures~\ref{fig:svm1} and ~\ref{fig:svm2}) with the best SVM configurations for all three feature sets.
Blue marks the results using the $S+H$ feature set, cyan the $H$ feature set, while yellow the $S$ feature set.
$S+H$ results are superior, while $H$ results are still better than $S$ results for all measures.
The chart shows that there is a constant 1-2\% improvement in all aspects of the SVM model results if we add the hybrid metrics to the feature sets.

\begin{figure}[htb!]
\begin{center}
\includegraphics[width=\textwidth]{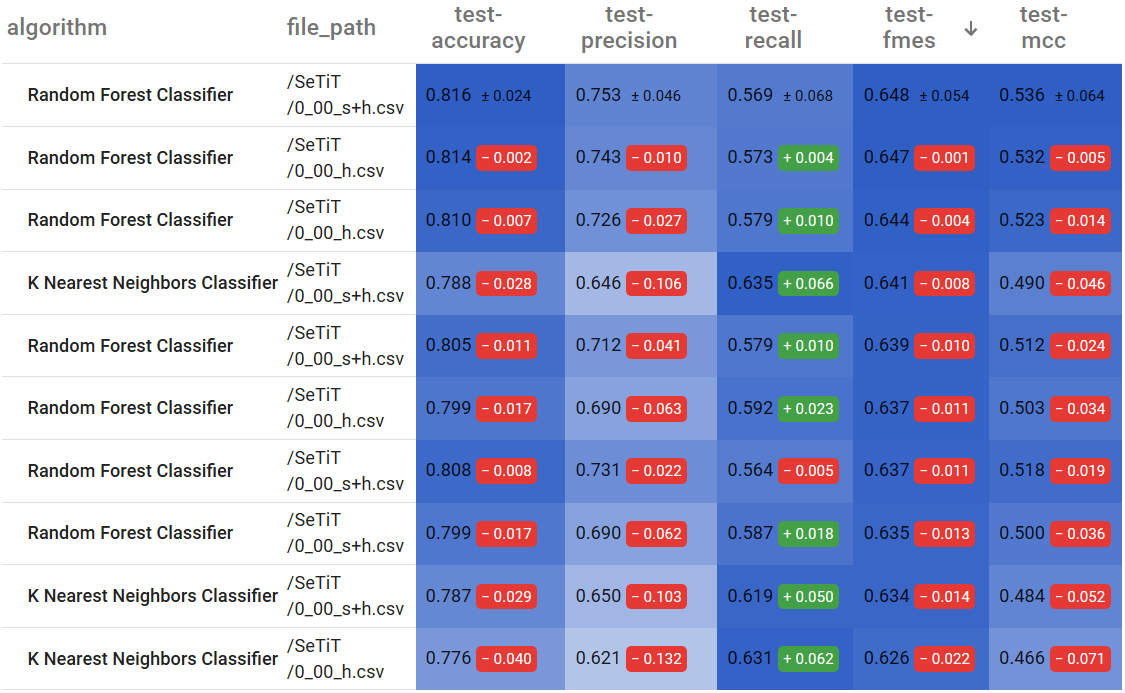}
\caption{Top 10 F-measures}
\label{fig:fmes}
\end{center}
\end{figure}

Figure~\ref{fig:fmes} displays a heat-mapped table of the top 10 model results based on their F-measure values.
We ranked all 108 models, meaning that all three feature sets are on the same list.
We can calculate F-measure with the following formula:
$$
    F-measure = 2\cdot\frac{Precision \cdot Recall}{Precision + Recall}.
$$
As we can see, Random Forest (0.648) and KNN (0.641) models achieve the best F-measures on the $S+H$ feature set.
The same Random Forest models produce almost as high F-measures (0.647) using only the $H$ feature set.
The best performing model on the $S$ feature set is not even in the top 10.
It shows that hybrid invocation metrics do increase the performance of ML models in terms of F-measure, meaning they improve the models' overall performance.
The best values are achieved by keeping both the original NOI and NII metrics and adding their hybrid counterparts HNOI and HNII, but using only the latter ones as substitutes for the static metrics still improves F-measure significantly.

\begin{figure}[htbp]
\begin{center}
\includegraphics[width=0.85\textwidth]{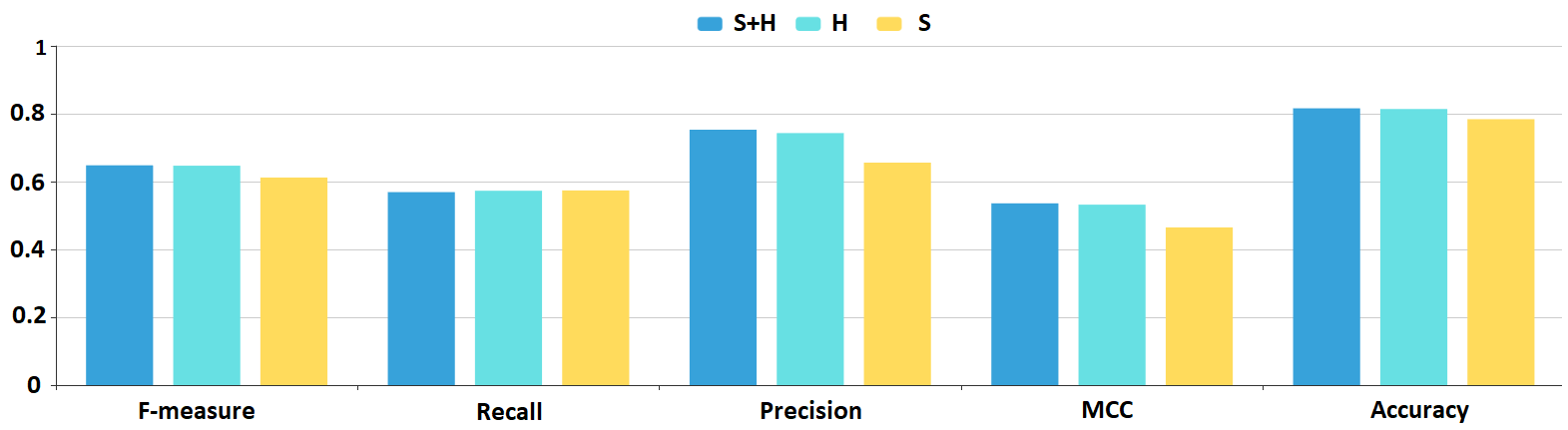}
\caption{Random forest}
\label{fig:rf1}
\end{center}
\end{figure}

\begin{figure}[htbp]
\begin{center}
\vspace{-20px}
\includegraphics[width=0.85\textwidth]{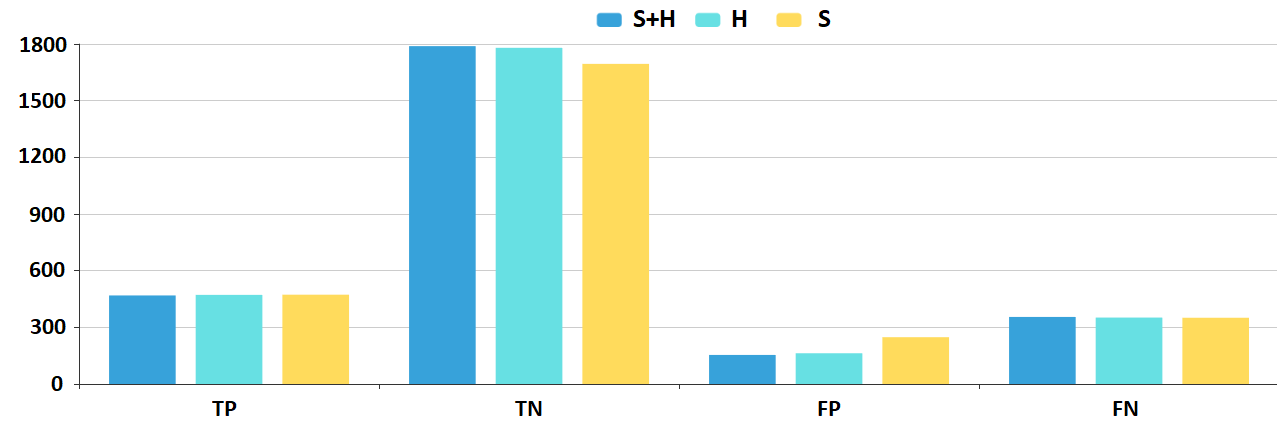}
\caption{Random forest}
\label{fig:rf2}
\end{center}
\end{figure}

To visualize the difference in the various performance measures, we plotted a bar-chart (Figures~\ref{fig:rf1} and ~\ref{fig:rf2}) with the best Random Forest configurations for all three feature sets.
Blue marks the results using the $S+H$ feature set, cyan the $H$ feature set, while yellow the $S$ feature set.
$S+H$ and $H$ results are better than $S$ results for all measures except for recall, but the difference there is only marginal.
The chart shows that there is a constant 1-2\% improvement in all aspects of the Random Forest model results, but precision is higher by approximately 10\% if we add the hybrid metrics to the feature sets.

\subsection{The Most Balanced Algorithm}

\begin{figure}[htbp]
\begin{center}
\includegraphics[width=0.85\textwidth]{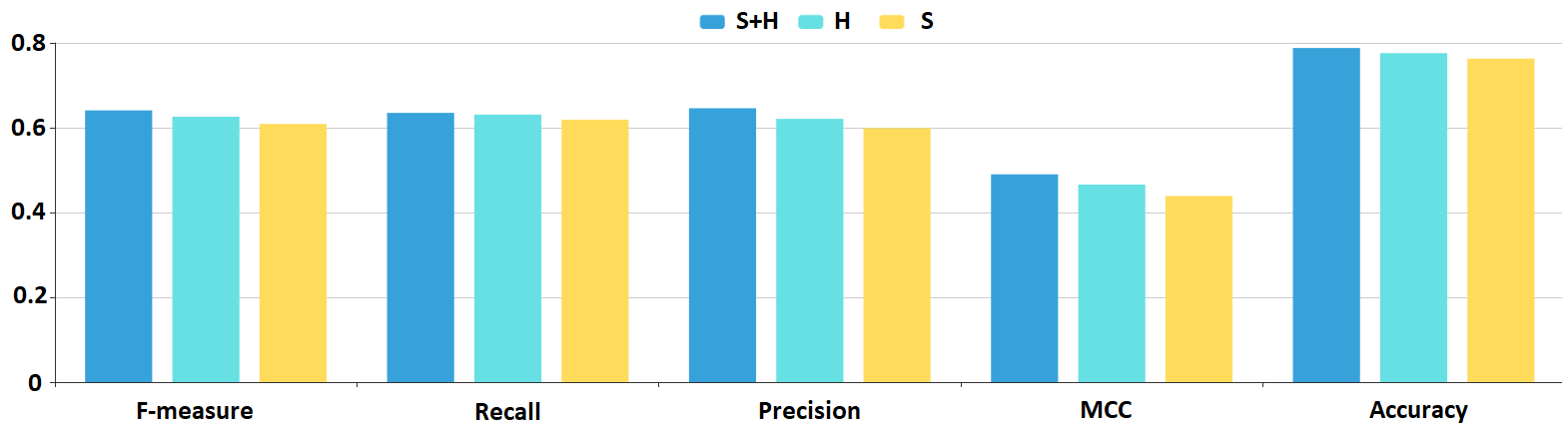}
\caption{K-nearest neighbors}
\label{fig:knn1}
\end{center}
\end{figure}

\begin{figure}[htbp]
\begin{center}
\includegraphics[width=0.85\textwidth]{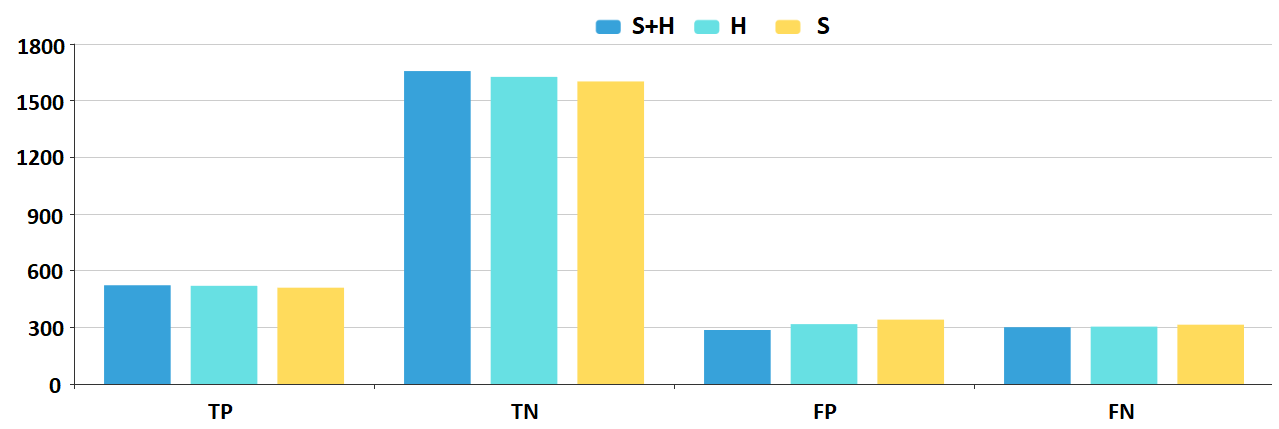}
\caption{K-nearest neighbors}
\vspace{-10px}
\label{fig:knn2}
\end{center}
\end{figure}

K-nearest neighbor models stand out in that they produce the most balanced performance measures.
As can bee sen in Figures~\ref{fig:knn1} and \ref{fig:knn2}, both precision and recall values are above 0.6, therefore F-measure is above 0.6 as well.
For this model, $H$ feature set brings a 1-2\% improvement over the $S$ feature set, while $S+H$ feature set results in a 2-5\% increase in performance.

\subsection{Significance Analysis of the Performance Measures}
Despite a seemingly consistent increase in every model performance measures caused by adding hybrid source code metrics to the features, we cannot be sure that this improvement is statistically significant.
Therefore, we performed a Wilcoxon signed-rank test~\cite{wilcoxon1970critical} on the model F-measure values between each pair of feature sets (S vs. H, S vs. S+H, H vs. S+H).
The detailed results (T statistics and p-values) are shown in Table~\ref{tab:f-wilcox}.

\begin{table}[htbp]
  \centering
  \caption{Wilcoxon signed-rank test results of F-measures between models using different feature sets}
    \begin{tabular}{llllll}
              & \multicolumn{2}{c}{\textbf{Hybrid}} & \multicolumn{2}{c}{\textbf{Static+Hybrid}}\\
     \textbf{Features} & \textbf{T} & \textbf{p-value} & \textbf{T} & \textbf{p-value} \\
    $>$ Static & \multicolumn{1}{r}{95.5} & \multicolumn{1}{r}{0.00032} & \multicolumn{1}{r}{74} & \multicolumn{1}{r}{0.00004}\\
    $>$ Hybrid & -     & -     & \multicolumn{1}{r}{116} & \multicolumn{1}{r}{0.0019} \\    
    \end{tabular}%
  \label{tab:f-wilcox}%
\end{table}%

As can be seen, the F-measure values achieved by the models differ significantly (p-value is less than 0.05) depending on the feature sets we used for training them.
It is interesting to observe that there is a significant difference in performances even between the models using the hybrid and static+hybrid features and not just between models using static features only and models using hybrid features as well.
These results confirm that even though hybrid source code metrics provide additional prediction power to bug prediction models, they do not substitute static source code metrics but complement them.
Therefore, according to our empirical data, we could achieve the best performing JavaScript bug prediction models by keeping static call-related metrics and adding their hybrid counterparts to the model features.
We note that the Wilcoxon signed-rank test showed significant results in the case of precision and recall performance measures as well.

\subsection{Results Overview and Discussion}

In the previous sections, we analyzed the best performing algorithms with a focus on the improvements caused by the hybrid source code metrics.
However, to have a complete picture of the results, we summarize the performances of all nine machine learning algorithms here.
Table~\ref{tab:comparison} shows the best prediction performances (i.e., models with best performing hyper-parameters and feature set) of all nine algorithms according to their F-measures.

\begin{table}[htbp]
  \centering
  \caption{The best results of the nine ML algorithms according to their F-measure}
    \begin{tabular}{llrrrrr}
    \textbf{ML algorithm} & \textbf{Feature set} & \multicolumn{1}{l}{\textbf{Accuracy}} & \multicolumn{1}{l}{\textbf{Precision}} & \multicolumn{1}{l}{\textbf{Recall}} & \multicolumn{1}{l}{\textbf{F-measure}} & \multicolumn{1}{l}{\textbf{MCC}} \\
    > Random Forest Classifier & S+H   & 0.816 & 0.753 & 0.569 & 0.648 & 0.54 \\
    > K Nearest Neighbors Classifier  & S+H   & 0.788 & 0.646 & 0.635 & 0.641 & 0.49 \\
    > Customized DNN Classifier & S+H   & 0.784 & 0.649 & 0.601 & 0.624 & 0.47 \\
    > Decision Tree Classifier & S+H   & 0.781 & 0.649 & 0.58  & 0.612 & 0.46 \\
    > Standard DNN Classifier & H     & 0.774 & 0.634 & 0.569 & 0.6   & 0.44 \\
    > Logistic Regression Classifier & S+H   & 0.787 & 0.682 & 0.533 & 0.598 & 0.46 \\
    > Support Vector Machine Classifier & S+H   & 0.789 & 0.699 & 0.515 & 0.593 & 0.47 \\
    > Linear Regression Classifier & S+H   & 0.769 & 0.67  & 0.443 & 0.533 & 0.4 \\
    > Naive Bayes Classifier & S+H   & 0.772 & 0.713 & 0.394 & 0.508 & 0.4 \\
    \end{tabular}%
  \label{tab:comparison}%
\end{table}%

There are several properties to observe in the table.
First, all but one ML model achieves the best results in terms of F-measure using the $S+H$ feature set.
The only exception is the Standard DNN Classifier, which performs best using only the hybrid version of call metrics (i.e., $H$ feature set).
Second, the Random Forest Classifier has the best F-measure ($0.648$) but also the best accuracy ($0.816$), precision ($0.753$), and MCC ($0.54$) metrics, which mean it performs the best overall for predicting software bugs in the studied JavaScript program.
However, there is a trade-off between precision and recall; therefore, the Random Forest Classifier has only the third-highest recall metric ($569$).
Third, the K Nearest Neighbors Classifier has the one but last lowest precision ($0.646$) and only the third-highest accuracy ($0.788$), still its recall ($0.635$) is the highest among all the models and in terms of F-measure ($0.641$) and MCC ($0.49$) it is a very close second behind the Random Forest Classifier.
It implicates that the K Nearest Neighbors Classifier achieves the most balanced performance, not the best in every aspect but very high performance measures with no large variance.
Fourth, the deep neural networks (Standard DNN Classifier and Customized DNN Classifier) do not outperform the simpler, classical models in this prediction task.
The most likely cause of this is the relatively small amount of training samples, so the real strength of deep learning cannot be exploited.
Fifth, the models struggle to achieve high recall values in general, which seems to be the bottleneck of the maximum F-measures.
Even the worst-performing models (Linear Regression Classifier and Naive Bayes Classifier) have an acceptable accuracy ($0.769$ and $0.772$, respectively) and precision ($0.67$ and $0.713$, respectively), but very low recall ($0.443$ and $0.394$, respectively), which results in a very low F-measure ($0.533$ and $0.508$, respectively).

To sum up our experiences, it is worthwhile to add hybrid call metrics to the set of standard static source code metrics for training a JavaScript bug prediction model.
To achieve the highest accuracy and precision, one should choose the Random Forest Classifier method, but if the recall is also important and one wants to have as balanced results as possible, the K Nearest Neighbors Classifier is the best possible option.

\section{Threats to Validity}\label{sec:threats}

There are several threats to the validity of the presented empirical study.
As a training set, we used 333 bugs only from one system.
Therefore, the results might be specific to this system and might not generalize well.
However, ESLint is a large and diverse program containing a representative set of issues.
Additionally, bugs are manually filtered, thus do not introduce noise in the prediction models.
As a result, we believe that our study is meaningful, though replication with more subject systems would be beneficial.

The threshold value chosen for calculating the hybrid call edges might affect the ML model performances.
We selected a threshold of $0$ (i.e., counted every edge with a weight greater than zero) in our case study; however, we carried out a sensitivity analysis with different thresholds as well.
Even though the calculated HNOI and HNII values changed based on the applied threshold, the model improvements using these values proved to be consistent with the ones presented in the study.
Therefore, we believe that the essence of the results is independent of the choice of the particular threshold value.

Finally, the provided thresholds might be inaccurate as we derived them from a manual evaluation of a small sample of real call edge candidates.
To eliminate the risk of human error, two senior researchers evaluated all the edges who had to agree on each call label.
For sampling, we applied a stratified selection strategy, so we evaluated more call samples from subsets of tools finding more edges in general, thus increasing the confidence of the derived weights.

\section{Conclusions}\label{sec:conclusion}

In this paper, we proposed a function level JavaScript bug prediction model based on static source code metrics with the addition of a hybrid (static and dynamic) code analysis based metrics for incoming and outgoing function calls.
JavaScript is a highly dynamic scripting language for which static code analysis might be very imprecise; therefore, combining static and dynamic analysis to extract features is a promising approach.

We created three versions of a training dataset from the functions of the ESLint project.
We used the BugsJS public dataset to find, extract, and map buggy functions in ESLint.
We ended up with a dataset containing 824 buggy and 1943 non-buggy functions with three sets of features: static metrics only, static metrics where the invocation metrics (NOI and NII) are replaced by their hybrid counterparts (HNOI and HNII), static metrics with the addition of the hybrid metrics.

We trained nine different models in 108 configurations and compared their results.
We found that using invocation metrics calculated by a hybrid code analysis as bug prediction features consistently improves the performance of the ML prediction models.
Depending on the ML algorithm, applied hyper-parameters, and target measure we consider, hybrid invocation metrics bring a 2-10\% increase in model performances (i.e., precision, recall, F-measure).
Interestingly, even though replacing static NOI and NII metrics with their hybrid counterparts HNOI and HNII in itself improve model performances, most of the time, keeping them both yields the best results.
It means that they hold somewhat complementary information to each other.
To achieve the highest accuracy and precision, one should choose the Random Forest Classifier method, but if the recall is also important and one wants to have as balanced results as possible, the K Nearest Neighbors Classifier is the best possible option.

\acknowledgments{
The presented work was carried out within the SETIT Project (2018-1.2.1-NKP-2018-00004)\footnote{Project no. 2018-1.2.1-NKP-2018-00004 has been implemented with the support provided from the National Research, Development and Innovation Fund of Hungary, financed under the 2018-1.2.1-NKP funding scheme.} and partially supported by grant TUDFO/47138-1/2019-ITM of the Ministry for Innovation and Technology, Hungary.
Furthermore, Péter Hegedűs was supported by the Bolyai János Scholarship of the Hungarian Academy of Sciences and the ÚNKP-20-5-SZTE-650 New National Excellence Program of the Ministry for Innovation and Technology.
}

\reftitle{References}
\bibliography{references}

\begin{thebibliography}{-------}
\providecommand{\natexlab}[1]{#1}

\bibitem[Gray(2013)]{gray2013software}
Gray, D.P.H.
\newblock Software defect prediction using static code metrics: formulating a
  methodology {\bf 2013}.

\bibitem[Li and Leung(2011)]{li2011mining}
Li, L.; Leung, H.
\newblock Mining static code metrics for a robust prediction of software
  defect-proneness.
\newblock  2011 International Symposium on Empirical Software Engineering and
  Measurement. IEEE,  2011, pp. 207--214.

\bibitem[Gray \em{et~al.}(2009)Gray, Bowes, Davey, Sun, and
  Christianson]{gray2009using}
Gray, D.; Bowes, D.; Davey, N.; Sun, Y.; Christianson, B.
\newblock Using the support vector machine as a classification method for
  software defect prediction with static code metrics.
\newblock  International Conference on Engineering Applications of Neural
  Networks. Springer,  2009, pp. 223--234.

\bibitem[Ferzund \em{et~al.}(2008)Ferzund, Ahsan, and
  Wotawa]{ferzund2008analysing}
Ferzund, J.; Ahsan, S.N.; Wotawa, F.
\newblock Analysing bug prediction capabilities of static code metrics in open
  source software. In {\em Software Process and Product Measurement}; Springer,
   2008; pp. 331--343.

\bibitem[Hata \em{et~al.}(2012)Hata, Mizuno, and Kikuno]{hata2012bug}
Hata, H.; Mizuno, O.; Kikuno, T.
\newblock Bug prediction based on fine-grained module histories.
\newblock  2012 34th international conference on software engineering (ICSE).
  IEEE,  2012, pp. 200--210.

\bibitem[Shivaji \em{et~al.}(2012)Shivaji, Whitehead, Akella, and
  Kim]{shivaji2012reducing}
Shivaji, S.; Whitehead, E.J.; Akella, R.; Kim, S.
\newblock Reducing features to improve code change-based bug prediction.
\newblock {\em IEEE Transactions on Software Engineering} {\bf 2012}, {\em
  39},~552--569.

\bibitem[Madeyski and Jureczko(2015)]{madeyski2015process}
Madeyski, L.; Jureczko, M.
\newblock Which process metrics can significantly improve defect prediction
  models? An empirical study.
\newblock {\em Software Quality Journal} {\bf 2015}, {\em 23},~393--422.

\bibitem[Binkley \em{et~al.}(2009)Binkley, Feild, Lawrie, and
  Pighin]{binkley2009increasing}
Binkley, D.; Feild, H.; Lawrie, D.; Pighin, M.
\newblock Increasing diversity: Natural language measures for software fault
  prediction.
\newblock {\em Journal of Systems and Software} {\bf 2009}, {\em
  82},~1793--1803.

\bibitem[Haiduc \em{et~al.}(2016)Haiduc, Arnaoudova, Marcus, and
  Antoniol]{haiduc2016use}
Haiduc, S.; Arnaoudova, V.; Marcus, A.; Antoniol, G.
\newblock The use of text retrieval and natural language processing in software
  engineering.
\newblock  Proceedings of the 38th International Conference on Software
  Engineering Companion,  2016, pp. 898--899.

\bibitem[Alshehri \em{et~al.}(2018)Alshehri, Goseva-Popstojanova, Dzielski, and
  Devine]{alshehri2018applying}
Alshehri, Y.A.; Goseva-Popstojanova, K.; Dzielski, D.G.; Devine, T.
\newblock Applying machine learning to predict software fault proneness using
  change metrics, static code metrics, and a combination of them.
\newblock  SoutheastCon 2018. IEEE,  2018, pp. 1--7.

\bibitem[Di~Nucci \em{et~al.}(2018)Di~Nucci, Palomba, De~Rosa, Bavota, Oliveto,
  and De~Lucia]{di2018developer}
Di~Nucci, D.; Palomba, F.; De~Rosa, G.; Bavota, G.; Oliveto, R.; De~Lucia, A.
\newblock A developer centered bug prediction model.
\newblock {\em IEEE Transactions on Software Engineering} {\bf 2018}, {\em
  44},~5--24.

\bibitem[Goyal \em{et~al.}(2013)Goyal, Chandra, and Singh]{goyal2013impact}
Goyal, R.; Chandra, P.; Singh, Y.
\newblock Impact of Interaction in the Combined Metrics Approach for Fault
  Prediction.
\newblock {\em Software Quality Professional} {\bf 2013}, {\em 15}.

\bibitem[Gyimesi \em{et~al.}(2019)Gyimesi, Vancsics, Stocco, Mazinanian,
  \'{A}rp\'{a}d Besz\'{e}des, Ferenc, and Mesbah]{2019-Gyimesi-ICST}
Gyimesi, P.; Vancsics, B.; Stocco, A.; Mazinanian, D.; \'{A}rp\'{a}d
  Besz\'{e}des.; Ferenc, R.; Mesbah, A.
\newblock {BugsJS}: a Benchmark of JavaScript Bugs.
\newblock  Proceedings of 12th IEEE International Conference on Software
  Testing, Verification and Validation (ICST),  2019, pp. 90--101.

\bibitem[F.(1974)]{allen1974interprocedural}
F., A.
\newblock {I}nterprocedural {D}ata {F}low {A}nalysis.
\newblock  Information Processing 74 (Software). North-Holland Publishing Co.,
  Amsterdam, The Netherlands,  1974, pp. 398--402.

\bibitem[{Ryder}(1979)]{ryder1979constructing}
{Ryder}, B.G.
\newblock Constructing the Call Graph of a Program.
\newblock {\em IEEE Transactions on Software Engineering} {\bf 1979}, {\em
  SE-5},~216--226.

\bibitem[Graham \em{et~al.}(1982)Graham, Kessler, and
  Mckusick]{Graham:1982:GCG:872726.806987}
Graham, S.L.; Kessler, P.B.; Mckusick, M.K.
\newblock {G}prof: {A} {C}all {G}raph {E}xecution {P}rofiler.
\newblock {\em SIGPLAN Not.} {\bf 1982}, {\em 17},~120--126.

\bibitem[Xie and Notkin(2002)]{xie2002empirical}
Xie, T.; Notkin, D.
\newblock {A}n {E}mpirical {S}tudy of {J}ava {D}ynamic {C}all {G}raph
  {E}xtractors.
\newblock {\em University of Washington CSE Technical Report 02-12} {\bf 2002},
  {\em 3}.

\bibitem[Murphy \em{et~al.}(1998)Murphy, Notkin, Griswold, and
  Lan]{murphy1998empirical}
Murphy, G.C.; Notkin, D.; Griswold, W.G.; Lan, E.S.
\newblock {{A}n {E}mpirical {S}tudy of {S}tatic {C}all {G}raph {E}xtractors}.
\newblock {\em ACM Trans. Softw. Eng. Methodol.} {\bf 1998}, {\em 7},~158--191.

\bibitem[Eichinger \em{et~al.}(2010)Eichinger, Pankratius, Gro{\ss}e, and
  B{\"o}hm]{eichinger2010localizing}
Eichinger, F.; Pankratius, V.; Gro{\ss}e, P.W.; B{\"o}hm, K.
\newblock {L}ocalizing {D}efects in {M}ultithreaded {P}rograms by {M}ining
  {D}ynamic {C}all {G}raphs. In {\em Testing--Practice and Research
  Techniques}; Springer,  2010; pp. 56--71.

\bibitem[Dmitriev(2004)]{dmitriev2004profiling}
Dmitriev, M.
\newblock {P}rofiling {J}ava {A}pplications {U}sing {C}ode {H}otswapping and
  {D}ynamic {C}all {G}raph {R}evelation.
\newblock {\em SIGSOFT Softw. Eng. Notes} {\bf 2004}, {\em 29},~139--150.

\bibitem[Eisenbarth \em{et~al.}(2001)Eisenbarth, Koschke, and
  Simon]{eisenbarth2001aiding}
Eisenbarth, T.; Koschke, R.; Simon, D.
\newblock {A}iding {P}rogram {C}omprehension by {S}tatic and {D}ynamic
  {F}eature {A}nalysis.
\newblock  Proceedings of the IEEE International Conference on Software
  Maintenance (ICSM'01). IEEE Computer Society,  2001, p. 602.

\bibitem[Feldthaus \em{et~al.}(2013)Feldthaus, Sch\"{a}fer, Sridharan, Dolby,
  and Tip]{feldthaus_acg}
Feldthaus, A.; Sch\"{a}fer, M.; Sridharan, M.; Dolby, J.; Tip, F.
\newblock {E}fficient {C}onstruction of {A}pproximate {C}all {G}raphs for
  {J}ava{S}cript {IDE} {S}ervices.
\newblock  Proceedings of the 2013 International Conference on Software
  Engineering; IEEE Press: Piscataway, NJ, USA,  2013; ICSE '13, pp. 752--761.

\bibitem[{Yu}(2019)]{yu2019empirical}
{Yu}, L.
\newblock Empirical Study of Python Call Graph.
\newblock  2019 34th IEEE/ACM International Conference on Automated Software
  Engineering (ASE),  2019, pp. 1274--1276.
\newblock
  doi:{\changeurlcolor{black}\href{https://doi.org/10.1109/ASE.2019.00160}{\detokenize{10.1109/ASE.2019.00160}}}.

\bibitem[Bolin(2010)]{bolin2010closure}
Bolin, M.
\newblock {\em Closure: {T}he {D}efinitive {G}uide: {G}oogle {T}ools to {A}dd
  {P}ower to {Y}our {J}ava{S}cript}; " O'Reilly Media, Inc.",  2010.

\bibitem[Fink and Dolby()]{fink2012wala}
Fink, S.; Dolby, J.
\newblock {WALA--The TJ Watson Libraries for Analysis}.

\bibitem[Antal \em{et~al.}(2018)Antal, Hegedus, T{\'o}th, Ferenc, and
  Gyim{\'o}thy]{antal2018static}
Antal, G.; Hegedus, P.; T{\'o}th, Z.; Ferenc, R.; Gyim{\'o}thy, T.
\newblock Static {J}ava{S}cript {C}all {G}raphs: {A} comparative study.
\newblock  2018 IEEE 18th International Working Conference on Source Code
  Analysis and Manipulation (SCAM). IEEE,  2018, pp. 177--186.

\bibitem[Wei and Ryder(2013)]{wei2013practical}
Wei, S.; Ryder, B.G.
\newblock Practical blended taint analysis for JavaScript.
\newblock  Proceedings of the 2013 International Symposium on Software Testing
  and Analysis. ACM,  2013, pp. 336--346.

\bibitem[Dijkstra(2014)]{dijkstra2014evaluation}
Dijkstra, J.
\newblock {E}valuation of {S}tatic {J}ava{S}cript {C}all {G}raph {A}lgorithms.
\newblock PhD thesis, Software Analysis and Transformation,  2014.

\bibitem[Madsen \em{et~al.}(2013)Madsen, Livshits, and
  Fanning]{madsen2013practical}
Madsen, M.; Livshits, B.; Fanning, M.
\newblock Practical {S}tatic {A}nalysis of {J}avaScript {A}pplications in the
  {P}resence of {F}rameworks and {L}ibraries.
\newblock  Proceedings of the 2013 9th Joint Meeting on Foundations of Software
  Engineering. ACM,  2013, pp. 499--509.

\bibitem[Lhot{\'a}k \em{et~al.}(2007)Lhot{\'a}k et~al.]{lhotak2007comparing}
Lhot{\'a}k, O.; others.
\newblock {C}omparing {C}all {G}raphs.
\newblock  Proceedings of the 7th ACM SIGPLAN-SIGSOFT workshop on Program
  analysis for software tools and engineering. ACM,  2007, pp. 37--42.

\bibitem[Ali and Lhot{\'a}k(2012)]{ali2012application}
Ali, K.; Lhot{\'a}k, O.
\newblock {A}pplication-{O}nly {C}all {G}raph {C}onstruction.
\newblock  ECOOP 2012 -- Object-Oriented Programming; Noble, J., Ed.; Springer
  Berlin Heidelberg: Berlin, Heidelberg,  2012; pp. 688--712.

\bibitem[Kinable and Kostakis(2011)]{kinable2011malware}
Kinable, J.; Kostakis, O.
\newblock {M}alware {C}lassification {B}ased on {C}all {G}raph {C}lustering.
\newblock {\em Journal in computer virology} {\bf 2011}, {\em 7},~233--245.

\bibitem[Eichinger \em{et~al.}(2008)Eichinger, B{\"o}hm, and
  Huber]{eichinger2008mining}
Eichinger, F.; B{\"o}hm, K.; Huber, M.
\newblock {{M}ining {E}dge-{W}eighted {C}all {G}raphs to {L}ocalise {S}oftware
  {B}ugs}.
\newblock  Machine Learning and Knowledge Discovery in Databases; Springer
  Berlin Heidelberg: Berlin, Heidelberg,  2008; pp. 333--348.

\bibitem[Rao and Steiner()]{rao2013debugging}
Rao, A.; Steiner, S.J.
\newblock {D}ebugging {F}rom a {C}all {G}raph.
\newblock US Patent 8,359,584.

\bibitem[Musco \em{et~al.}(2016)Musco, Monperrus, and Preux]{Musco2016Large}
Musco, V.; Monperrus, M.; Preux, P.
\newblock A Large Scale Study of Call Graph-based Impact Prediction using
  Mutation Testing.
\newblock {\em Software Quality Journal} {\bf 2016}, {\em 25}.
\newblock
  doi:{\changeurlcolor{black}\href{https://doi.org/10.1007/s11219-016-9332-8}{\detokenize{10.1007/s11219-016-9332-8}}}.

\bibitem[Munaiah and Meneely(2016)]{Munaiah2016BeyondTA}
Munaiah, N.; Meneely, A.
\newblock Beyond the Attack Surface: Assessing Security Risk with Random Walks
  on Call Graphs.
\newblock  SPRO '16,  2016.

\bibitem[Nguyen and Tran(2010)]{nguyen2010predicting}
Nguyen, V.H.; Tran, L.M.S.
\newblock Predicting Vulnerable Software Components with Dependency Graphs.
\newblock  Proceedings of the 6th International Workshop on Security
  Measurements and Metrics; Association for Computing Machinery: New York, NY,
  USA,  2010; MetriSec '10.
\newblock
  doi:{\changeurlcolor{black}\href{https://doi.org/10.1145/1853919.1853923}{\detokenize{10.1145/1853919.1853923}}}.

\bibitem[{Cheng} \em{et~al.}(2019){Cheng}, {Wang}, {Hua}, {Zhang}, {Xu}, {Yi},
  and {Sui}]{cheng2019static}
{Cheng}, X.; {Wang}, H.; {Hua}, J.; {Zhang}, M.; {Xu}, G.; {Yi}, L.; {Sui}, Y.
\newblock Static Detection of Control-Flow-Related Vulnerabilities Using Graph
  Embedding.
\newblock  2019 24th International Conference on Engineering of Complex
  Computer Systems (ICECCS),  2019, pp. 41--50.

\bibitem[Neuhaus \em{et~al.}(2007)Neuhaus, Zimmermann, Holler, and
  Zeller]{neuhaus2007predicting}
Neuhaus, S.; Zimmermann, T.; Holler, C.; Zeller, A.
\newblock Predicting Vulnerable Software Components.
\newblock  2007, pp. 529--540.
\newblock
  doi:{\changeurlcolor{black}\href{https://doi.org/10.1145/1315245.1315311}{\detokenize{10.1145/1315245.1315311}}}.

\bibitem[Lee \em{et~al.}(2010)Lee, Jeong, and Lee]{Lee2010Detecting}
Lee, J.; Jeong, K.; Lee, H.
\newblock Detecting Metamorphic Malwares Using Code Graphs.
\newblock  Proceedings of the 2010 ACM Symposium on Applied Computing;
  Association for Computing Machinery: New York, NY, USA,  2010; SAC '10, p.
  1970–1977.
\newblock
  doi:{\changeurlcolor{black}\href{https://doi.org/10.1145/1774088.1774505}{\detokenize{10.1145/1774088.1774505}}}.

\bibitem[Punia \em{et~al.}(2014)Punia, Kumar, and Sharma]{punia2014evaluation}
Punia, S.K.; Kumar, A.; Sharma, A.
\newblock Evaluation the quality of software design by call graph based
  metrics.
\newblock {\em Global Journal of Computer Science and Technology} {\bf 2014}.

\bibitem[Just \em{et~al.}(2014)Just, Jalali, and Ernst]{JustJE2014}
Just, R.; Jalali, D.; Ernst, M.D.
\newblock {Defects4J}: A {Database} of existing faults to enable controlled
  testing studies for {Java} programs.
\newblock  ISSTA 2014, Proceedings of the 2014 International Symposium on
  Software Testing and Analysis; ,  2014; pp. 437--440.
\newblock Tool demo.

\bibitem[Antal \em{et~al.}(2020)Antal, Tóth, Hegedűs, and
  Ferenc]{antal2020enhancedTrainingData}
Antal, G.; Tóth, Z.G.; Hegedűs, P.; Ferenc, R.
\newblock {Enhanced Bug Prediction in JavaScript Programs with Hybrid
  Call-Graph Based Invocation Metrics (Training Dataset)},  2020.
\newblock
  doi:{\changeurlcolor{black}\href{https://doi.org/10.5281/zenodo.4281476}{\detokenize{10.5281/zenodo.4281476}}}.

\bibitem[Ferenc \em{et~al.}(2020)Ferenc, Viszkok, Aladics, Jász, and
  Hegedűs]{FERENC2020100551}
Ferenc, R.; Viszkok, T.; Aladics, T.; Jász, J.; Hegedűs, P.
\newblock Deep-water framework: The Swiss army knife of humans working with
  machine learning models.
\newblock {\em SoftwareX} {\bf 2020}, {\em 12},~100551.
\newblock
  doi:{\changeurlcolor{black}\href{https://doi.org/https://doi.org/10.1016/j.softx.2020.100551}{\detokenize{https://doi.org/10.1016/j.softx.2020.100551}}}.

\bibitem[Wilcoxon \em{et~al.}(1970)Wilcoxon, Katti, and
  Wilcox]{wilcoxon1970critical}
Wilcoxon, F.; Katti, S.; Wilcox, R.A.
\newblock Critical values and probability levels for the Wilcoxon rank sum test
  and the Wilcoxon signed rank test.
\newblock {\em Selected tables in mathematical statistics} {\bf 1970}, {\em
  1},~171--259.

\end{thebibliography}

\end{document}